# Some comments on Wojcik's hypothesis

S. Piekarski, J. Wojcik

(Institute of Fundamental Technological Research Polish Academy of Sciences, Warsaw Poland)

Wojcik's hypothesis has been mentioned on page 32 of the text S. Piekarski, "Galilean – Invariant Formulation of the Fluid Mechanics", IFTR REPORTS, 7/2007. Here we discuss it in more detail. Our main aim is to show that the form of the restrictions imposed by the Gibbs identity can depend on the choice of coordinates. The possible reactions for this unpleasant fact are shortly discussed.

## I. Introduction

In the monograph "Linear and Nonlinear Waves" G.B. Whitham writes on p.151, under **6.4 Thermodynamic Relations**:
"…We could just take the view that $e(p,\rho)$ in (6.24) is some empirical function that is given to us. However, the arguments developed in thermodynamics not only provide us with formulas but suggest the important quantities to consider. It is appropriate to note here only the mathematical steps we require, and refer the reader to numerous standard texts for motivation and a study of the deeper significance of the issues.
    The differential form

$$de + pd\left(\frac{1}{\rho}\right) \qquad (6.30)$$

plays a fundamental role. It arises first in considering the consequences when a small amount of energy is added to unit mass of the gas. If the energy is added relatively slowly so that there is no violent change in the pressure, the work done in expanding the volume $\frac{1}{\rho}$ by $pd\left(\frac{1}{\rho}\right)$ is $pd\left(\frac{1}{\rho}\right)$. The rest of the energy must go into increasing the internal energy by $de$. In these circumstances (6.30) is equal to the amount of energy added. But in any event, for given $e(p,\rho)$ it is a differential form in two variables $p, \rho$.
    By Pfaff's theorem, this always has an integrating factor, so that there exists functions $T(p,\rho)$ and $S(p,\rho)$ such that

$$TdS = de + pd\left(\frac{1}{\rho}\right) \qquad (6.31)$$

This simple mathematical statement acquires its deep significance from the fact that $T$ is the absolute temperature and $S$ is the entropy.
    In more complicated systems, other thermodynamic variables (such as the concentration of different phases of the substance) appear besides $p$ and $\rho$.

Then the differential form corresponding to (6.30) involves more than two variables. On purely mathematical grounds one can no longer claim that there is always an integrating factor to relate it to a perfect differential. However, the basis of thermodynamics is that this will always be so for all real physical systems, and, moreover, the integrating factor will always be the absolute temperature.

The mathematical step in (6.31) seems to introduce $T$ and $S$ as subsidiary derived quantities from a given $e(p,\rho)$. But they play an equally fundamental role, and (6.31) should be viewed more as a relation between equally important quantities.

*Ideal Gas.*

Under normal conditions most gases obey the ideal gas law

$$p = R\rho T \tag{6.32}$$

where $R$ is a constant. When that is the case, we can write (6.31) as

$$dS = \frac{de}{T} - d(R\log\rho).$$

It follows that $\frac{de}{T}$ must be a perfect differential and therefore $e$ is a function of $T$ alone:

$$e = e(T). \tag{6.33}$$

It is interesting that (6.33) can be deduced from the assumption (6.32), but actually (6.33) is more fundamental.
Equations 6.32 and 6.33 describe an ideal gas. In the equations of motion it is convenient to express $e$ as function of $p$ and $\rho$. We see that for an ideal gas $e$ is a function of $p/\rho$. The form of this function could be left open, but in fact a rather simple formula covers a wide range of phenomena in gas dynamics. It arises in considerations of the specific heats...."

In the above text the notation and numeration of [1] has been applied.

It is worth to discuss Whitham's text in more detail and that shall be done in the next chapter. Conclusions are at the end of this paper.

## II. Basic Formulas

In the notation applied in [2] an ideal gas is explicitly defined as

$$E = AT, \tag{2.1}$$

$$p = B\rho T, \tag{2.2}$$

where $T$ denotes the temperature, $\rho$ is the mass density, $p$ is the pressure, $E$ denotes the energy density per unit mass.

Let

$$de + pd\left(\frac{1}{\rho}\right) \tag{2.3}$$

be a one – form on a two – dimensional manifold $Q$, written, in general, in arbitrary coordinates from a complete atlas on $Q$.

Pfaff's theorem states that (2.3) has an integrating factor and therefore the following identity is satisfied

$$d\left\{\Phi\left[de + pd\left(\frac{1}{\rho}\right)\right]\right\} = 0. \tag{2.4}$$

It can be easily checked that the inverse of the temperature $T$ taken from (2.1) and (2.2) is an integrating factor for the 1 – form

$$dAT + B\rho Td\left(\frac{1}{\rho}\right), \tag{2.5}$$

that is, a differential of the 1 – form

$$\frac{1}{T}\left[dAT + B\rho Td\left(\frac{1}{\rho}\right)\right] \tag{2.6}$$

vanishes:

$$d\left\{\frac{1}{T}\left[dAT + B\rho Td\left(\frac{1}{\rho}\right)\right]\right\} = d\left\{\left[\frac{1}{T}dAT + \frac{1}{T}B\rho Td\left(\frac{1}{\rho}\right)\right]\right\} =$$

$$d\left\{\left[\frac{A}{T}dT + B\rho d\left(\frac{1}{\rho}\right)\right]\right\} = d\left\{\left[\frac{A}{T}dT + B\rho\left[-\frac{1}{\rho^2}\right]d\rho\right]\right\} =$$

$$d\left\{\left[\frac{A}{T}dT - \frac{B}{\rho}d\rho\right]\right\} = d\left[\frac{A}{T}\right] \wedge dT - d\left[\frac{B}{\rho}\right] \wedge d\rho =$$

$$-\frac{A}{T^2}dT \wedge dT + \frac{B}{\rho^2}d\rho \wedge d\rho = 0. \tag{2.7}$$

Now one can ask what expressions for energy density are allowed if the energy density is defined in the variables $\rho, T$, the integrating factor is the inverse of the temperature $T$ and the expression for the pressure is of the form (2.2).

In other words, the allowed expression for the energy $E(\rho, T)$ must satisfy the identity

$$d\left\{\frac{1}{T}\left[dE(\rho,T) + B\rho Td\left(\frac{1}{\rho}\right)\right]\right\} =$$

$$d\left\{\frac{1}{T}dE(\rho,T) + B\rho d\left(\frac{1}{\rho}\right)\right\} = 0. \tag{2.8}$$

Now, any function $E(\rho,T)$ can be decomposed in the form

$$E(\rho,T) = AT + E'(\rho,T). \tag{2.9}$$

After inserting (2.9) into (2.8) one obtains

$$d\left\{\frac{1}{T}dE(\rho,T) + B\rho d\left(\frac{1}{\rho}\right)\right\} =$$
$$d\left\{\frac{1}{T}d[AT + E'(\rho,T)] + B\rho d\left(\frac{1}{\rho}\right)\right\} = 0. \tag{2.10}$$

Obviously,

$$d\left\{\frac{1}{T}d[AT + E'(\rho,T)] + B\rho d\left(\frac{1}{\rho}\right)\right\} =$$
$$d\left\{\frac{1}{T}d[AT] + B\rho d\left(\frac{1}{\rho}\right)\right\} + d\left\{\frac{1}{T}d[E'(\rho,T)]\right\} = 0. \tag{2.11}$$

However, from (2.7) we already know that the first term in the second line of (2.11) is identically equal to zero and therefore we are left with the condition

$$d\left\{\frac{1}{T}d[E'(\rho,T)]\right\} = 0. \tag{2.12}$$

The condition (2.13) shows the restrictions imposed on the allowed forms of the energy functions by the Gibbs identity (expressed in coordinates $\rho,T$). The restrictions of that kind are discussed in more detail in [2].

Next one can change the variables $\rho,T$ to the variables $\rho,p$ (possibility of such a transformation is a direct consequence of (2.2)): from

$$p = B\rho T$$

one can determine the temperature $T$ as a function of $p$ and $\rho$ (with $B$ as a parameter):

$$T = \frac{p}{B\rho} \tag{2.13}$$

Then the energy density for an ideal gas takes the form

$$AT = \frac{Ap}{B\rho}.$$
(2.14)

The interesting fact is that (according to Whitham's scheme) in these variables one does not obtain any restrictions on the possible forms of the energy function.
In the identity

$$d\left\{\Phi\left[dE(\rho,p) + pd\left(\frac{1}{\rho}\right)\right]\right\} = 0$$
(2.15)

The temperature can be defined as the inverse of the integrating factor and on account of Pfaff's theorem such procedure does not impose any restrictions on the energy function. Therefore, we are forced to conclude that the form of the restrictions imposed by the Gibbs identity (according to the procedure described by Whitham) depends on the choice of coordinates.

## III. Conclusions

The above example seems to suggest that in modeling the behavior of a fluid one encounters the following alternative.
The first possibility is to fix the coordinates in which the Gibbs identity is written (as a part of a model of a fluid).
The second possibility is to formulate the model of a fluid in a manner independent on the choice of coordinates, that is, according to the Wojcik's hypothesis [2].
Of course, a detailed comparison between both options is outside the scope of this text. Here, we wanted only to initiate a discussion on possible inconsistencies in Whitham's presentation [1].